\documentclass[12pt]{article}
\usepackage{oldgerm}
\usepackage{euler}
\usepackage{graphics}
\usepackage{bm}
\usepackage{graphicx}
\usepackage{amsmath}
\usepackage{amssymb}
\usepackage{amstext}
\usepackage{amscd}
\usepackage{amsfonts}
\usepackage{appendix}
\usepackage{color}
\DeclareMathAlphabet{\EuFrak}{U}{euf}{m}{n}
\DeclareMathAlphabet{\EuScript}{U}{eus}{m}{n}

\newcommand{\nd}{\noindent}

\newcommand{\be}{\begin{equation}}
\newcommand{\ee}{\end{equation}}
\newcommand{\ben}{\begin{eqnarray}}
\newcommand{\een}{\end{eqnarray}}

\title{{\bf Hypergeometric foundations of  Fokker-Plank like equations}}

\author{{A. Plastino$^1$, M.C.Rocca$^1$} \\
\small{$^1$ La Plata National University and
Argentina's National Research Council}\\
\small{(IFLP-CCT-CONICET)-C. C. 727, 1900 La Plata - Argentina}}
\date{\today}

\begin{document}

\maketitle

\begin{abstract}

We show that the Fokker Planck equation   can  be derived from a
Hypergeometric differential equation. The same applies to a non
linear generalization of such equation.\vskip 3mm

\nd {\bf Keywords:} Non linear Fokker-Plank equations,
separation of variables, hypergeometric function.

\end{abstract}

\newpage

\renewcommand{\theequation}{\arabic{section}.\arabic{equation}}

\section{Introduction}

\nd In this paper we uncover the  fact that the celebrated
Fokker-Planck (FP) equation \cite{risken}

\be \label{fp} \frac{\partial F}{\partial t}=
-\frac{\partial}{\partial x} [K(x) F]  + \frac{Q}{2}\,
\frac{\partial^2 F}{\partial x^2}, \ee can be derived from an
hypergeometric differential equation.  In this equation, $F$ is
the distribution function, $K(x)$ the drift coefficient and $Q$
the diffusion coefficient (a positive quantity) \cite{risken}. The
second term on the r.h.s, describes the effects of the fluctuating
forces (diffusion term). Without it, (\ref{fp}) would describe
deterministic motion (over-damped motion of a particle under the
force $K(x)$).  For the time being, we restrict ourselves to the
case $K={\rm constant}$. A similar hypergeometric derivation
applies to {\bf a non linear generalization} of  equation
(\ref{fp}), in the spirit if the one advanced 20 years ago by
Plastino and Plastino \cite{ppFP}, that has arisen interest till
today \cite{ii1,ii2,ii3,ii4}.

\nd This papers continues a line of research initiated by
uncovering hypergeometric connotations of quantum equations
\cite{tp1}.

\setcounter{equation}{0}

\section{Hypergeometric derivation of the Fokker-Plank equation}
The ordinary hypergeometric function $F_1^2(a,b;c;z)$ is a special
function represented by the hypergeometric series, that includes
many other special functions as specific or limiting cases.  It is
a solution of a second-order linear ordinary differential equation
(ODE). Many second-order linear ODEs can be transformed into this
equation. Generalized hypergeometric functions include the
confluent hypergeometric function as a special case, which in turn
have many particular special functions as special instances, such
as elementary functions, Bessel functions, and the classical
orthogonal polynomials.

\nd In particular, the confluent hypergeometric function reads
\cite{tp1}\vskip 3mm

\ben \label{conflu}  &  \phi(a,b,z)=
\sum\limits_{n=0}^\infty\,\frac{a_n}{b_n}\,\frac{z^n}{n!}; \,\,a,\,b \in
{\mathcal{R}}, \cr\cr & {\rm with \,\,a_n, \,\,b_n \,\,the
\,\,Pochhamer \,\,symbols}\cr\cr &
 a_0=1, a_n=a\,(a+1)\,(a+2)\,...\,(a+n-1);\,\,{\rm same\,\,for}\,\,b.\een

\nd The confluent hypergeometric equation satisfies the
differential equation \cite{tp1}:
\begin{equation}
\label{eq2.1}
z\phi^{''}(a,b,z)+(b-z)\phi^{'}(a,b,z)-a\phi(a,b,z)=0,
\end{equation}
that, for  $a=b$ adopts the appearance
\begin{equation}
\label{eq2.2}
z\phi^{''}(a,a,z)+(a-z)\phi^{'}(a,a,z)-a\phi(a,a,z)=0.
\end{equation}
Accordingly,
\begin{equation}
\label{eq2.3} \phi(a,a,z)=e^z.
\end{equation}
Let us consider now the function $\phi$ and the variable $\lambda$
\begin{equation}
\label{eq2.4} \phi\left[a,a,-\left(\lambda t+\frac {x}
{\lambda}\right)\right]= e^{-\left(\lambda t+\frac {x}
{\lambda}\right)},
\end{equation}
where we express  $\lambda$ in terms of an equation involving two
quantities $K$ and $Q$ of Eq. (\ref{fp})
\begin{equation}
\label{eq2.5} \lambda^3+K\lambda+\frac {Q} {2}=0,
\end{equation}
and we define $z$ as
\begin{equation}
\label{eq2.6} z=-\lambda t-\frac {x} {\lambda}.
\end{equation}
Given that  $\phi$ is such that
\begin{equation}
\label{eq2.7} \phi^{''}=\lambda^2\frac {\partial^2\phi} {\partial
x^2}\;\;\; ;\;\;\; \phi^{'}=-\frac {1} {\lambda}\frac
{\partial\phi} {\partial t} \equiv \phi,
\end{equation}
Eq. (\ref{eq2.2}) can be recast as
\begin{equation}
\label{eq2.8} z\lambda^2\frac {\partial^2\phi} {\partial
x^2}+a\phi^{'} +\frac {z} {\lambda}\frac {\partial\phi} {\partial
t}-a\phi=0.
\end{equation}
Since $\phi^{'}=\phi$, (\ref{eq2.8}) gets simplified to

\begin{equation}
\label{eq2.9} \lambda^3\frac {\partial^2\phi} {\partial x^2}
+\frac {\partial\phi} {\partial t}=0.
\end{equation}
According to  (\ref{eq2.5}), Eq. (\ref{eq2.9}) becomes
\begin{equation}
\label{eq2.10} -(K\lambda+\frac {Q} {2})\frac {\partial^2\phi}
{\partial x^2} +\frac {\partial\phi} {\partial t}=0.
\end{equation}
In addition, since $\phi$ verifies
\begin{equation}
\label{eq2.11} \lambda\frac {\partial^2\phi} {\partial x^2}=
-\frac {\partial\phi} {\partial x},
\end{equation}
we are led to the following expression for  (\ref{eq2.10})
\begin{equation}
\label{eq2.12} K\frac {\partial\phi} {\partial x} -\frac {Q}
{2}\frac {\partial^2\phi} {\partial x^2} +\frac {\partial\phi}
{\partial t}=0,
\end{equation}
which is tantamount to
\begin{equation}
\label{eq2.13} \frac {\partial\phi} {\partial t}+ \frac {\partial
(K\phi)} {\partial x} -\frac {Q} {2}\frac {\partial^2\phi}
{\partial x^2}=0,
\end{equation}
  i.e., Fokker-Plank's equation  para $K$ independent of $x$. Of course, when $K$ does depend upon $x$
one just {\it postulates}  (\ref{eq2.13}). Note that, by
definition,  (\ref{eq2.4}) is a solution of  (\ref{eq2.13}).

\setcounter{equation}{0}

\section{Separation of variables in the Ornstein-Uhlenbeck process $K=x$}

\nd The Ornstein–Uhlenbeck process  is a stochastic process that,
loosely, describes the velocity of a massive Brownian particle
under the influence of friction. It is stationary, Gaussian, and
Markovian, being the only nontrivial evolution that satisfies
these three conditions, up to allowing for linear transformations
of the space and time variables. We believe that this well known
process of linear drift \cite{risken} is worth revisiting for
didactic purposes. We start with
\begin{equation}
\label{eq3.1} \frac {\partial F} {\partial t}+ \frac {\partial
(KF)} {\partial x} -\frac {Q} {2}\frac {\partial^2F} {\partial
x^2}=0,
\end{equation}
\begin{equation}
\label{eq3.2} F(x,t)=G(t)H(x),
\end{equation}
which leads to
\begin{equation}
\label{eq3.3} \frac {1} {G}\frac {\partial G} {\partial t}= \frac
{1} {H}\left[ \frac {Q} {2}\frac {\partial^2H} {\partial x^2}
-\frac {\partial (KH)} {\partial x}\right]=-\lambda,
\end{equation}
with $\lambda>0$. From here we are immediately led to
\begin{equation}
\label{eq3.4} \frac {\partial G} {\partial t}+\lambda G=0,
\end{equation}
\begin{equation}
\label{eq3.5} \frac {Q} {2}\frac {d^2H} {d x^2} -\frac {d (KH)} {d
x}+\lambda H=0.
\end{equation}
For the linear instance  $K=-x$ we first obtain for $G$
\begin{equation}
\label{eq3.6} G(t)=e^{-\lambda t}.
\end{equation}
Applying the Fourier transform to (\ref{eq3.5}) we find
\begin{equation}
\label{eq3.7} \frac {Q} {2}\alpha^2\hat{H}+ \alpha\frac {d\hat{H}}
{d\alpha}-\lambda\hat{H}=0,
\end{equation}
where  $\hat{H}$ is the  Fourier transform of $H$ of variable
$\alpha$. One solves (\ref{eq3.7}) and get
\begin{equation}
\label{eq3.8} \hat{H}(\alpha)=|\alpha|^{\lambda}e^{-\frac
{Q\alpha^2} {4}},
\end{equation}
and from  (\ref{eq3.8}) we encounter for  $H$
\begin{equation}
\label{eq3.9} H(x)=\frac {1} {2\pi}\int\limits_{-\infty}^{\infty}
|\alpha|^{\lambda}e^{-\frac {Q\alpha^2} {4}} e^{-i\alpha
x}\;d\alpha.
\end{equation}
Thus we have for $F$ the general expression
\begin{equation}
\label{eq3.10} F(x,t)=\frac {1} {2\pi}
\int\limits_0^{\infty}\int\limits_{-\infty}^{\infty} \lambda
a(\lambda)e^{-\lambda t} |\alpha|^{\lambda}e^{-\frac {Q\alpha^2}
{4}} e^{-i\alpha x}\;d\alpha\;d\lambda,
\end{equation}
where  $a(\lambda)$ must verify
\begin{equation}
\label{eq3.11} \int\limits_0^{\infty} a(\lambda)\;d\lambda=1.
\end{equation}
Eq. (\ref{eq3.10}) may have been obtained before, but we were
unable to find such derivation in the vast FP-literature available
to us. \setcounter{equation}{0}

\section{Non-Linear Fokker-Plank Equation \cite{book}}

\nd Anomalous diffusion is exhibited in a variety of physical
systems and is therefore the subject of much interest. It can be
observed, for example, in general systems such as plasma flow,
porous media, and surface growth, as well as in more specific
situations such as cytltrimethylammonium bromide miscelles
dissolved in salted water and NMR relaxometry of liquids in porous
glasses \cite{book}. The main characteristic of anomalous
diffusion is the fact that the mean squared displacement is not
proportional to time $t$ but rather to some power of it. If the
scaling is faster than $t$, then  the pertinent system is
superdiffusive while, if it is slower than $t$,  it is
subdiffusive. A  nonlinear Fokker-Planck diffusion equation has
been proposed for those systems with correlated anomalous
diffusion, beginning with \cite{ppFP} and followed afterward by,
for instance, \cite{otros,lisa,nobre}. For an excellent  overview,
see \cite{book}.

\nd For the ordinary hypergeometric function $F_1^2(a,b;c;z)$ we
have \cite{tp2}, using now three Pochhamer symbols,

\be F_1^2(a,b;c;z) \equiv F(a,b;c;z)= \sum_{n=0}^\infty\,
\frac{a_{(n)}b_{(n)}}{c_{(n)}}\,\frac{z^n}{n!};\,\,(|z| < 1 ),
\label{gral} \ee where the series terminates if either $a$ or $b$
is a non-zero integer. A particularly important special case is

\be \label{binomio} F(-m,b,b,-z) = (1+z)^m.   \ee Eq. (\ref{gral})
verifies  \cite{tp2}

\begin{equation}
\label{eq4.1} z(1-z)F^{''}(\alpha,\beta;\gamma;z)+
[\gamma-(\alpha+\beta+1)z]F^{'}(\alpha,\beta;\gamma;z)-
\alpha\beta F(\alpha,\beta;\gamma;z)=0.
\end{equation}
If $\beta=\gamma$, then  $F$ satisfies \cite{tp3}
\begin{equation}
\label{eq4.2} F(-\alpha,\gamma;\gamma;-z)=(1+z)^{\alpha}.
\end{equation}
Focus attention now upon the function
\begin{equation}
\label{eq4.3} f(x,t)=\left[1+(q-1)\left(\lambda t+\frac {x}
{\lambda} \right)\right]^{\frac {1} {1-q}},
\end{equation}
where  $\lambda$ obeys (for $K$ and $Q$ both constants)
\begin{equation}
\label{eq4.4} \lambda^3+K\lambda+\frac {Q} {2}=0.
\end{equation}
Recourse to  (\ref{eq4.2}) allows one to write
\begin{equation}
\label{eq4.5} F\left[\frac {1} {q-1},\gamma;\gamma;
(1-q)\left(\lambda t+\frac {x} {\lambda} \right)\right]=
\left[1+(q-1)\left(\lambda t+\frac {x} {\lambda}\right)
\right]^{\frac {1} {1-q}},
\end{equation}
and then
\begin{equation}
\label{eq4.6} z=(1-q)\left(\lambda t+\frac {x} {\lambda}\right).
\end{equation}
For $\beta=\gamma$, $F$ [Cf. (\ref{eq4.1})] adopts the appearance

\begin{equation}
\label{eq4.7} z(1-z)F^{''}(\alpha,\gamma;\gamma;z)+
[\gamma-(\alpha+\gamma+1)z]F^{'}(\alpha,\gamma;\gamma;z)-
\alpha\beta F(\alpha,\gamma;\gamma;z)=0.
\end{equation}
Since  $F$ verifies
\begin{equation}
\label{eq4.8} F^{''}=\frac {\lambda^2} {(1-q)^2}\frac {\partial^2
F} {\partial x^2}\;\;\;;\;\;\;F^{'}=\frac {1} {\lambda} \frac
{\partial F} {\partial t},
\end{equation}
then  (\ref{eq4.7}) becomes
\begin{equation}
\label{eq4.9} z(1-z)\frac {\lambda^2} {(1-q)^2}\frac {\partial^2
F} {\partial x^2}+ \frac {qz} {\lambda(1-q)^2} \frac {\partial F}
{\partial t}+\gamma(1-z)F^{'}- \frac {\gamma} {q-1}F=0,
\end{equation}
and, adequately simplifying,
\begin{equation}
\label{eq4.10} (1-z)\lambda^3\frac {\partial^2 F} {\partial x^2}+
q\frac {\partial F} {\partial t}+ \frac {\gamma} {z}(1-q)^2
\left[(1-z)F^{'}-\frac {1} {q-1}F\right]=0.
\end{equation}
Again, since  $F$ fulfills
\begin{equation}
\label{eq4.11}
(1-z)F^{'}-\frac {1} {q-1}F=0
\end{equation}
Eq. (\ref{eq4.10}) becomes
\begin{equation}
\label{eq4.12} (1-z)\lambda^3\frac {\partial^2 F} {\partial x^2}+
q\frac {\partial F} {\partial t}=0,
\end{equation}
or,  equivalently,
\begin{equation}
\label{eq4.13} \lambda^3F^{(1-q)}\frac {\partial^2 F} {\partial
x^2}+ q\frac {\partial F} {\partial t}=0,
\end{equation}
since $F^{(1-q)}(z)=1-z$. Thus, we are in a position to cast
(\ref{eq4.13}) as
\begin{equation}
\label{eq4.14} \lambda^3\frac {\partial^2 F} {\partial x^2}+ \frac
{\partial F^q} {\partial t}=0.
\end{equation}
Utilizing (\ref{eq4.4}) we can recast things as

\begin{equation}
\label{eq4.15} -\left(\lambda K+\frac {Q} {2}\right)\frac
{\partial^2 F} {\partial x^2}+ \frac {\partial F^q} {\partial
t}=0.
\end{equation}
Remembering that  $F$ obeys
\begin{equation}
\label{eq4.16} \lambda K\frac {\partial^2 F} {\partial
x^2}=-K\frac {\partial F^q} {\partial x}= -\frac {\partial (KF^q)}
{\partial x},
\end{equation}
we obtain from  (\ref{eq4.15})
\begin{equation}
\label{eq4.17} \frac {\partial F^q} {\partial t}+ \frac
{\partial(KF^q)} {\partial x} -\frac {Q} {2}\frac {\partial^2 F}
{\partial x^2}=0,
\end{equation}
a nonlinear Fokker-Planck equation. We postulate its validity for
 $K=K(x)$ as well. If we set
 \begin{itemize}

\item $g = F^{q}$

\item $2-q^*= 1/q$,

\end{itemize}
we immediately ascertain that Eq. (\ref{eq4.17}), expressed in
terms of $g$ and $q^*$, coincides with the nonlinear FP postulated
by Plastino and Plastino in \cite{ppFP}

\be   \label{eq4.18} \frac {\partial g} {\partial t}  + \frac
{\partial(Kg)} {\partial x} -    \frac {Q} {2}\frac {\partial^2 g^{2-q^*}}
{\partial x^2}=0.   \ee

\vskip 3mm
 \nd For the stationary case ($F$ independent of
$t$) we have for Eq. (\ref{eq4.17})
\begin{equation}
\label{eq4.19} \frac {\partial(K(x) F^q)} {\partial x} -\frac {Q}
{2}\frac {\partial^2 F} {\partial x^2}=0,
\end{equation}
whose solution is
\begin{equation}
\label{eq4.20} F(x)=\left[1+\frac {2(q-1)} {Q}V(x)\right]^{\frac
{1} {1-q}},
\end{equation}
where $\frac {dV(x)} {dx}=-K(x)$.

\setcounter{equation}{0}

\section{Conclusions}

\nd We have shown that the Fokker-Planck equation  and its
nonlinear generalization by Plastino and Plastino \cite{ppFP} are
contained within the structure  of hypergeometric  linear
differential equations, for constant drift $K$. The FP-extensions
to general drifts $K(x)$ have to be postulated like in the
ordinary cases.

\nd We have displayed a general solution for the Orstein-Uhlenbeck
equation of constant drift that possibly might be new, although we
cannot ascertain it.

\nd We also give an exact solution of the nonlinear FP equation
when $F$ does not depend upon the time.

\end{document}